# Identification of 3.55 KeV line in the framework of standard physics.

## Dubrovich V.K.

*St. Petersburg Branch of Special Astrophysical Observatory, St. Petersburg*

Identification of the X-ray 3.55 KeV line as a recombination line of $\pi^-$ tritium mesoatom is proposed. It has been shown that in principle it is possible to form such an atom under standard laboratory conditions, without bringing in new physics.

At the moment there is a lot of data on continuum and spectral soft X-ray observation of various astrophysical objects. Such type of research is well developed and it allows obtaining interesting and important information on the sources. Among results there may be some unexpected, difficult to explain, and hypothesis-generating ones. The most important discoveries are those that may need new laws of physics to be explained by. The discovery of X-ray spectral line with energy equal to 3.55 KeV is a particular example (E. Bulbul et al., 2014). Analysis made with the help of standard catalogues hasn't shown and correspondence of such line to any known highly ionized ions. Currently the situation around identification of the line is so complicated that it has given rise to speculations regarding nonstandard physics involved in its origination (Cline, et al., 2014, Colon, et al., 2014). Thereupon thorough investigation of ways of the line identification, provided by standard physics, is of current importance.

Here one possible way, based on well-known physical phenomena, is suggested. Line spectrum of mesoatom of tritium and helium where electron is substituted by $\pi^-$ meson is considered. It is shown that there is a possibility of appearance of such an atom under standard astrophysical conditions. Energy of one of the energy levels differences coincides with measured energy of 3.55 KeV.

In considered interpretation, closeness of measured 3.55 KeV energy of photons to the energy of Lyman continuum quantum ($L_c$) in atom, consisting of proton $p$ and negative $\pi^-$ meson, plays the key role (see, e.g., D. F. Anagnostopoulos, et al., 2001). Given classical formula for hydrogen-like atom, taking into account meson and proton mass ratio, one can obtain for the energy $E_n$ of the level $n$ (Landau, Lifshitz, 1974):

$$E_n = \frac{Z^2 m_\pi e^4}{2\hbar^2(1+\frac{m_\pi}{M})}\frac{1}{n^2} = \alpha^2 Z^2 \frac{m_\pi c^2}{2(1+\frac{m_\pi}{M})n^2} \qquad (1)$$

where $Z$ – nucleus charge measured in units of electron charge, $m_\pi$ – mass of $\pi$ meson, $e$ – electron charge, $\hbar$ – the Plank constant, $M$ – nucleus mass, $c$ – the

speed of light, $\alpha = 1/137.036\$$ – the fine-structure constant. For $m_\pi c^2 = 139.569$ MeV and $M= m\_p=938.28$ MeV (1.0078 a.e.) one obtains:

$$E_n = 3.235 \text{ KeV} /n^2 \qquad (2)$$

Energy of line $L_c$ corresponds here to $n=1$. The difference between this quantity and the measured one (taking into account measurement accuracy ~30 eV as in Cline, et al., 2014) obviously put an obstacle in the way of needed identification. On the other hand, there is one more circumstance, which apparently makes attempts to form such mesoatom under standard conditions practically impossible, – short lifetime of $\pi$ meson – $\tau = 2.6 \cdot 10^{-8}$ seconds. Consider recombination of $\pi^-$ meson onto a proton, so the energy of free meson should be of order of line width – 30 eV. Then its velocity **v** should be **v** $\approx 10^7$ cm/sec. Taking into account the lifetime of meson, distance to the proton where meson is born should be less than 0.3 cm! That is meson should be born in the immediate vicinity of the nucleus. In laboratory experiments on mesoatoms mesons are born in a gas, dense enough for them to cool down quickly (D. F. Anagnostopoulos, at al., 2001). In astrophysical conditions, especially in regions, where there is only very rare gas in galaxy clusters, such mechanism of cooling isn't the case. So we need to consider reactions of $\pi^-$ birth in cases of high-energy particles scattering on the same nucleus, where further recombination of meson takes place. At the same time we shouldn't restrict ourselves by only free proton.

Consider some birth reactions of $\pi^-$ mesons with low energy:

1. $p + \gamma = p + \pi^+ + \pi^-$
2. $p + e^- = \Lambda + \nu_e \; ; \; \Lambda = p + \pi^-$
3. $n + e^- = \Sigma + \nu_e \; ; \; \Sigma = n + \pi^-$
4. $p + p = p + p + \pi^+ + \pi^-$

Here the following notations are used: $p$ – proton, $\gamma$ – gamma-quantum, $e^-$ – electron, $\Lambda$ – $\Lambda$ hyperon, $\nu_e$ – electron neutrino, $n$ – neutron, $\Sigma$ – $\Sigma$ hyperon. Within the limits of energy and momentum conservation laws, the first reaction allows to obtain a close pair $p + \pi^-$ with relative velocities of particles in pair close to zero, which causes recombination in mesoatom. However, initial energy of gamma-quantum required is rather high $E_\gamma > 300$ MeV, and the newborn mesoatom has very high velocity, which significantly broadens the recombination line. The second reaction, due to energy and momentum conservation laws doesn't allow the formation of proton and meson with very close velocities, so it doesn't lead to mesoatom as the first reaction does. Therefore scattering of an electron on a free proton does not fit to our purposes. However, we may obtain meson with almost

zero momentum is such reaction. Indeed, from energy and momentum conservation laws, taking momentum of meson equal to zero, one can deduce:

$$\mathbf{p}_\Lambda = \mathbf{p}_p = \mathbf{p} \qquad (3)$$

$$M_\Lambda c^2 + \mathbf{p}^2/2M_\Lambda = m_p c^2 + m_\pi c^2 + \mathbf{p}^2/2m_p \qquad (4)$$

Here $\mathbf{p}_p$ – momentum of the resulting proton. We obtain:

$$|\mathbf{p}| = [2\, m^*\,(M_\Lambda c^2 - m_p c^2 - m_\pi c^2)]^{1/2} \qquad (5)$$

$$m^* = M_\Lambda m_p/(M_\Lambda + m_p) \qquad (6)$$

Thus we can see that reaction with zero momentum of meson in final state is possible in principle. For such reaction to happen it is necessary for the momentum of intermediately born Λ hyperon to be close to $\mathbf{p}$. The difference in rest energies of particles at the beginning and at the end of the reaction, enclosed in round parentheses, is $\Delta E = 37.6$ MeV, and $m^* = 0.55 m_p$. Velocity of the final proton is $\mathbf{v}_p = |\mathbf{p}|/m_p = 0.21c = 6.3\ 10^9$ cm/sec.

Now we can build up the scheme of the process we are interested in. Consider fast electron scattering on a helium atom $^4$He. Interaction with one of the protons according to second reaction leads to formation of slow meson and fast proton, which moves away from the nucleus with high speed. At the end we obtain tritium nucleus T and a cold meson. Using energy and momentum conservation laws applied to second reaction one can easily obtain estimation of electron's minimum energy $E_0$ (which brings to momentum of Λ hyperon $|\mathbf{p}|$) in (5): $E_0 \approx 210$ MeV. In case of such energy of incident electron, bond energy of proton in nucleus can be neglected. It follows from observations of cosmic rays that electron energy required isn't extraordinary. Meson, which forms afterwards, may have very low energy and furthermore it can have velocity vector directed towards the nucleus. As a result, it jumps to a lower energy level, while emitting a photon with energy close to $L_c$. Tritium nucleus mass T – $m_T = 2808$ MeV (3.0156 a.e.). We substitute that value in (1) and obtain:

$$E_\gamma = 3.540\ \text{KeV}$$

If meson has energy greater than some value (5-10 eV ?), then it gets in a free state with high probability and cannot recombine back during its lifetime.

In order to have more detailed picture in the case of mesoatoms it is necessary to take into account correction to the energy of ground state due to strong interaction between the meson and the nucleus. To do that in first approximation it is sufficient to estimate time fraction of meson being in effective area of nuclear

forces and multiply it by typical energy of nuclear interaction. For a meson in ground 1S state such fraction is approximately equal to the ratio of cubes of the nucleus radius $R_я$ (1.5÷2 $10^{-13}$ cm) and mesoatom Bohr radius $R_Б$ (≈2$10^{-11}$ cm). Typical bond energy of nucleons in the nucleus is of the order of 8 MeV per nucleon. Thus, the correction we are interested in, taking into account that there are three nucleons, is of order of 10 ÷ 20 eV. Finally we obtain:

$$E_\gamma = 3.55 \div 3.56 \text{ KeV} \qquad (8)$$

Analogous scheme can be realized in case of third reaction. Only in that case at first electron collides with neutron in nucleus of helium $^4$He. The output consists of fast neutron, slow meson and nucleus $^3$He with charge Z=2. The line we are interested in corresponds to Balmer limit $H_c - n=2$. However here we should obtain two more lines – $L_\alpha(=3H_c=10.65$ KeV$)$, $L_c$ $(=4H_c=14.2$ KeV$)$. The line 10.65 KeV can be identified with CuXXVIII – 10.63 KeV, while the line 14.2 KeV can be identified with Sr – 14.16 KeV. Thus relation of all three lines may be left unnoted after their identification. Strictly speaking, the same schemes should be present for $\mu^-$ meson (mass $m_\mu$ = 105.66 MeV). This decay channel of Λ hyperon is less probable by approximately four orders of magnitude than $\pi^-$ channel. Simple calculations performed in (1) gives $E_\gamma$ =2.711 KeV for tritium, while for $^3$He – 2.711 KeV, 8.133 KeV and 10.844 KeV.

All arguments and estimates made above only refer to the amount of energy of appearing X-ray quanta. Another problem is to determine intensity of the given radiation. To solve such problem one need to know dependence of cross sections for the second and third reactions on energies of the electron and the resulting meson. Also it is very important to calculate probability W of meson recombination on the nucleus correctly. Here it is necessary to take into consideration that Λ hyperon, which forms at the beginning, has almost the same velocity as the resulting proton, and thus during its lifetime $\tau_A$=2.6 $10^{-10}$ sec it moves away from the nucleus at the distance $r_0$~1 cm. Probability of recombination $W \sim \Omega \approx \sigma/(4\pi r^2)$, where σ – cross section of meson recombination, while r – distance between nucleus and the point where meson is born. As a matter of principle r can be much less than $r_0$ because, according to quantum mechanics, some decay reactions of Λ hyperons can go by the time $\tau << \tau_A$. Relative amount of such reactions $\delta \approx \tau/\tau_A$ – is a small quantity. Taking into account $r=c\tau=r_0\delta$, the overall amount of recombinations $N \sim W\delta \approx \sigma/(4\pi\delta r_0^2) \sim 1/\delta$. Comprehensive calculation of line intensity obviously should include information on parameters of the source. However, body of small parameters, characterizing precision of the process setting, causes very low resulting intensity. Therefore, it seems that it is observable only in large objects such as galaxy clusters with very hot electron gas. On the other hand, data on line intensity and its profile gives opportunity to estimate gas parameters of an astrophysical object.

The main disadvantage of schemes based on reactions 2 and 3 is their low probability.  This is due to as a small cross section of  Λ hyperon birth because of the weak interaction in the channel of reaction, as well as additional severe restrictions on the moment of its decay. The channel of reactions 1 and 4 may represent an alternative. Here there is no weak interaction and one can in principle obtain a cold $\pi^-$ meson. The same scheme then follows – helium transforms into tritium, two protons and $\pi^+$ obtain high velocities, while cold $\pi^-$ meson recombines on the tritium nucleus.  In contrast to the reactions 2 and 3 here initially we must have an energetic proton instead of an electron. However, the energy of the proton should be of the order of 300 MeV, which is rather possible in relativistic objects.

# Conclusion

It is shown that a cold $\pi^-$ meson, cold tritium nucleus ($^3$H) and fast particles (proton, meson, neutrino) can be born in process of collision of highly energetic particle (photon, electron, proton) with helium nucleus ($^4$He). Tritium nucleus captures cold $\pi^-$ meson and forms mesoatom with emission of a photon with energy of 3.55 KeV. The overall efficiency of different options cannot be estimated a priori, because it depends on many different factors.

It is significant that in the scheme of consideration a thin X-ray line appear in matter without heavy metals – only helium and energetic electrons participate in the process. In principal that rises up expectations of presence of such lines in spectra of very distant objects with primordial chemical composition.

Photon energy calculation using current scheme can be done also in case of electrons scattering on any nuclei heavier than helium, which can be of interest for dense, rich with heavy metals, atmospheres of stars and relativistic objects. The same process can be of interest for applications in laboratory conditions on Earth.